\documentclass[twocolumn,nofootinbib,amsmath,amssymb,a4paper]{revtex4}

\usepackage{graphicx}% Include figure files
\usepackage{dcolumn}% Align table columns on decimal point
\usepackage{bm}

\newcommand{\Btaun}{{B_u \to \tau \nu}}
\newcommand{\cB}{{\cal B}}

\newcommand{\beq}{\begin{equation}}
\newcommand{\eeq}{\end{equation}}
\newcommand{\beqa}{\begin{eqnarray}}
\newcommand{\eeqa}{\end{eqnarray}}

\newcommand{\cO}{{\cal O}}

\newcommand{\sla}[1]%
        {\kern .25em\raise.18ex\hbox{$/$}\kern-.75em #1}% Feynman slash
\newcommand{\mybar}[1]%
        {\kern 0.8pt\overline{\kern -0.8pt#1\kern -0.8pt}\kern 0.8pt}

\begin{document}

\title{Flavor Physics in SUSY at large $\tan\beta$}
\author{Paride Paradisi
%\footnote{Electronic address: chsmith@itp.unibe.ch}
}
\affiliation{Departament de F\'{\i}sica Te\`orica and IFIC, Universitat de 
Val\`encia-CSIC, E-46100, Burjassot, Spain.}

\begin{abstract}

We discuss the phenomenological impact of a particularly interesting
corner of the MSSM: the large $\tan\beta$ regime.
The capabilities of leptonic and hadronic Flavor Violating processes
in shedding light on physics beyond the Standard Model are reviewed.
Moreover, we show that tests of Lepton Universality in charged current
processes can represent an interesting handle to obtain relevant
information on New Physics scenarios.

\end{abstract}

\maketitle

\section{Introduction}

Despite the great phenomenological success of the Standard Model (SM),
it is natural to consider this theory only as the low-energy limit
of a more general model.

The direct exploration of New Physics (NP) particles at the TeV scale
will be performed at the upcoming LHC.
A complementary strategy in looking for NP is provided
by high-precision low-energy experiments where NP could be detected
through the virtual effects of NP particles.
In particular, flavor-changing neutral-current (FCNC) transitions
may exhibit a sensitivity reach even beyond that achievable by
the direct searches at the LHC while representing, at the same time,
the best (or even the only) tool to extract information about
the flavor structures of NP theories.

In view of the above considerations, it is clear that flavor physics provides 
necessary and complementary information to those obtainable by the LHC.

Besides FCNC decays, also the Lepton Flavor Universality (LFU) tests
($K_{\ell 2}$ and $\pi_{\ell 2}$) offer a unique opportunity to probe
the SM and thus, to shed light on NP: the smallness of NP effects is
more than compensated by the excellent experimental resolution and
the good theoretical control.
\section{LFV in SUSY}

The discovery of neutrino masses and oscillations has unambiguously 
pointed out the existence of the Lepton Flavor Violation (LFV) thus, 
we expect this phenomenon to occur also in the charged-lepton sector.

Within a SM framework with massive neutrinos, FCNC transitions in the
lepton sector like $\ell_i\to\ell_j\gamma$ are strongly suppressed by
the GIM mechanism at the level of
${\cB}(\ell_i\to\ell_j\gamma)\sim (m_{\nu}/m_{W})^{4}\sim 10^{-50}$
well beyond any realistic experimental resolution \cite{epxLFV}.
In this sense, the search for FCNC transitions of charged leptons is one
of the most promising directions where to look for physics beyond the SM.

Within a SUSY framework, LFV effects originate from any misalignment
between fermion and sfermion mass eigenstates.
In particular, if the light neutrino masses are obtained via a
see-saw mechanism, the radiatively induced LFV entries in the slepton
mass matrix $(m^2_{\tilde{L}})_{ij}$ are given by \cite{BorzumatiMasiero}:
\begin{equation}
\label{mLfromseesaw}
(m^2_{\tilde{L}})_{i\neq j} \approx
- \frac{3m^2_0}{8\pi^2} (Y_{\nu} Y_{\nu}^\dagger)_{i\neq j}
\ln \left(\frac{M_X}{M_{R}} \right)\,,
\end{equation}
where $M_X$ denote the scale of SUSY-breaking mediation and $m_0$ 
the universal supersymmetry breaking scalar mass.
Since the see--saw equation
\footnote{The effective light-neutrino mass matrix obtained from a see-saw
mechanism is $m_\nu = - Y_\nu \hat{M}^{-1}_R Y_\nu^T \langle H_u \rangle^2$,
where $\hat{M}_R$ is the $3\times 3$ right-handed neutrino mass matrix and
$Y_{\nu}$ are the $3\times 3$ Yukawa couplings between left- and right-handed
neutrinos (the potentially large sources of LFV), and $\langle H_u \rangle$ is
the vacuum expectation value of the up-type Higgs.}
allows large $(Y_\nu Y_\nu^\dagger)$ entries, sizable effects can stem from
this running \cite{BorzumatiMasiero}.

The determination of $(m^2_{\tilde{L}})_{i\neq j}$ 
would imply a complete knowledge of the neutrino Yukawa matrix
$(Y_\nu)_{ij}$, which is not possible even if all
the low-energy observables from the neutrino sector were known.
As a result, the predictions of leptonic FCNC effects will remain
undetermined even in the very optimistic situation where all the 
relevant NP masses were measured at the LHC.

This is in contrast with the quark  sector, where similar RGE contributions
are completely determined in terms of quark masses and  CKM-matrix elements.

More stable predictions can be obtained embedding the SUSY model within
a Grand Unified Theory (GUT) where the see-saw mechanism can naturally
arise (such as $SO(10)$). In this case the GUT symmetry allows us
to obtain some hints about the unknown neutrino Yukawa matrix $Y_{\nu}$.
Moreover, in GUT scenarios there are other contributions
stemming from the quark sector \cite{BarbieriHall}.
These effects are completely independent from the structure of $Y_{\nu}$ and
can be regarded as new irreducible LFV contributions within SUSY GUTs.
For instance, within $SU(5)$, as both $Q$ and $e^c$ are hosted in the
{\bf 10} representation, the CKM matrix mixing the left handed quarks
will give rise to off diagonal entries in the running of the right-handed
slepton soft masses \cite{BarbieriHall}.

There exist to different classes of LFV contributions to rare decays:
\begin{enumerate}

\item[i)] Gauge-mediated LFV effects through the exchange of gauginos and sleptons,
\item[ii)] Higgs-mediated LFV effects through effective non-holomorphic 
Yukawa interactions \cite{bkl}\,.
\end{enumerate}
The above contributions decouple with the heaviest mass in the
slepton/gaugino loops $m_{SUSY}$ (case $i)$) or with the heavy Higgs 
mass $m_H$ (case $ii)$).

In principle, $m_H$ and $m_{SUSY}$ refers to different mass scales.
Higgs mediated effects start being competitive with the gaugino 
mediated ones when $m_{SUSY}$ is roughly one order of
magnitude heavier then $m_H$ and for $\tan\beta\sim\mathcal O(50)$
\cite{paradisiH}.

%\subsection{Correlations among LFV transitions}

While the appearance of LFV transitions would unambiguously signal
the presence of NP, the underlying theory generating LFV 
phenomena will remain undetermined, in general.

A powerful tool to disentangle among NP theories is the study of
the correlations of LFV transitions among same families
\cite{RossiBrignole,paradisiH,buraslfv}.

Interestingly enough, the predictions for the correlations among
LFV processes are very different in the gauge- and Higgs-mediated cases
\cite{paradisiH}.
In this way, if several LFV transitions are observed, their correlated
analysis could shed light on the underlying mechanism of LFV.
In the case of gauge-mediated LFV amplitudes
the $\ell_{i}\rightarrow\ell_{j}\ell_{k}\ell_{k}$ decays
are dominated by the $\ell_{i}\rightarrow\ell_{j}\gamma^{*}$
dipole transition, which leads to the unambiguous prediction:
\beq
\frac{\cB(\ell_{i}\rightarrow \ell_{j}\ell_{k}\ell_{k})}{\cB(\ell_{i}\rightarrow \ell_{j}\gamma)}
\simeq \frac{\alpha_{el}}{3\pi}\left(\log\frac{m^2_{l_{i}}}{m^2_{l_{k}}}-3\right)
\eeq
\beq
\label{relations}
\frac{\cB(\mu - e {\rm\ in \ Ti})}{\cB(\mu\!\rightarrow\!e\gamma)}
\simeq\! 
\alpha_{el}\,.
\eeq
If some ratios different from the above were discovered, then this would be 
clear evidence that some new process is generating the $\ell_i\rightarrow \ell_j$ 
transition, with Higgs mediation being a potential candidate
\footnote{As recently shown in \cite{buraslfv}, a powerful tool to disentangle
between Little Higgs models with T parity (LHT) and SUSY theories is a
correlated analysis of LFV processes.
In fact, LHT and SUSY theories predict very different
correlations among LFV transitions \cite{buraslfv}.}.

As regards the Higgs mediated case, $Br(\tau\rightarrow l_j\gamma)$ still gets 
generally the largest contribution among all the possible LFV decay modes 
\cite{paradisiH}. The following approximate relations hold \cite{paradisiH}:
\beq
\frac{Br(\tau\rightarrow l_j\gamma)}{Br(\tau\rightarrow l_j\eta)}\gtrsim 1
\,,\qquad\frac{Br(\tau\rightarrow l_j\eta)}{Br(\tau\rightarrow l_j\mu\mu)} 
\gtrsim
\frac{36}{3\!+\!5\delta_{j\mu}}\,.
\eeq
\beq
\frac{Br(\tau\rightarrow l_j ee)}{Br(\tau\rightarrow l_j\mu\mu)}
\gtrsim\frac{0.4}{3\!+\!5\delta_{j\mu}}\,.
\eeq
\beqa
\frac{Br(\mu\rightarrow e \gamma)}{Br(\mu Al\rightarrow e Al)} &\sim& 
10\,,\qquad\frac{Br(\mu\rightarrow eee)}{Br(\mu \rightarrow e \gamma)}\sim\alpha_{el}\,.
\eeqa
On the other hand, a correlated study of processes 
of the same type but relative to different family transitions,
like $Br(\mu\rightarrow e\gamma)/Br(\tau\rightarrow\mu\gamma)
\sim [(m^2_{\tilde{L}})_{21}/(m^2_{\tilde{L}})_{32}]^2$,
provides important information about the unknown structure
of the LFV source, i.e. $(m^2_{\tilde{L}})_{i\neq j}$.
\section{LFU in SUSY}

High precision electroweak tests, such as deviations from the SM expectations
of the LFU breaking, represent a powerful tool to probe the SM and, hence,
to constrain or obtain indirect hints of new physics beyond it.
Kaon and pion physics are obvious grounds where to perform such tests, for instance 
in the $\pi\rightarrow\ell\nu_{\ell}$ and $K\rightarrow\ell\nu_{\ell}$ decays,
where $l= e$ or $\mu$. In particular, the ratios
\beq
R_P^{\mu/e} = \frac{ \cB(P\to \mu \nu) }{ \cB(P \to e \nu)}
\eeq
can be predicted with excellent accuracies in the SM, 
both for $P=\pi$ (0.02\% accuracy \cite{Marciano}) and $P=K$ 
(0.04\% accuracy \cite{Marciano}), allowing for some 
of the most significant tests of LFU.

As recently pointed out in Ref.~\cite{kl2}, large departures from the SM 
expectations can be generated within a SUSY framework with R-parity
only once we assume i) LFV effects, ii) large $\tan\beta$ values.

Denoting by $\Delta r^{e-\mu}_{\!NP}$ the deviation from $\mu-e$ 
universality in $R_{K}$ due to NP, i.e.:
$R_{K}^{\mu/e}=(R_{K}^{\mu/e})_{SM}\left(1+\Delta r^{e-\mu}_{K}\right)$,
it turns out that \cite{kl2}:
\begin{equation}
\label{lfv}
\Delta r^{e-\mu}_{K}
\simeq 
\left(\frac{m^{4}_{K}}{M^{4}_{H}}\right)
\!\left(\frac{m^{2}_{\tau}}{m^{2}_{e}}\right)|\Delta^{31}_{R}|^2\,
\tan^{\!6}\!\beta.
\end{equation}
The deviations from the SM could reach $\sim 1\%$ in the $R_K^{\mu/e}$ case
\cite{kl2} (not far from the present experimental resolution \cite{kl2_exp})
and $\sim {\rm few} \times 10^{-4}$ in the $R_\pi^{\mu/e}$ case while 
maintaining LFV effects in $\tau$ decays at the $10^{-10}$ level.
In the pion case the effect is quite below the present experimental
resolution~\cite{pl2_exp}, but could well be within the reach of the
new generation of high-precision $\pi_{\ell 2}$ experiments planned at
TRIUMPH and at PSI. Larger violations of LFU are expected in $B\to\ell\nu$
decays, with $\cO(10\%)$ deviations from the SM in $R_B^{\mu/\tau}$ and
even order-of-magnitude enhancements in $R_B^{e/\tau}$~\cite{tgbhints}.
\section{Flavor physics at large $\tan\beta$ and Dark Matter}

\begin{figure*}[h]
\includegraphics[scale=0.4]{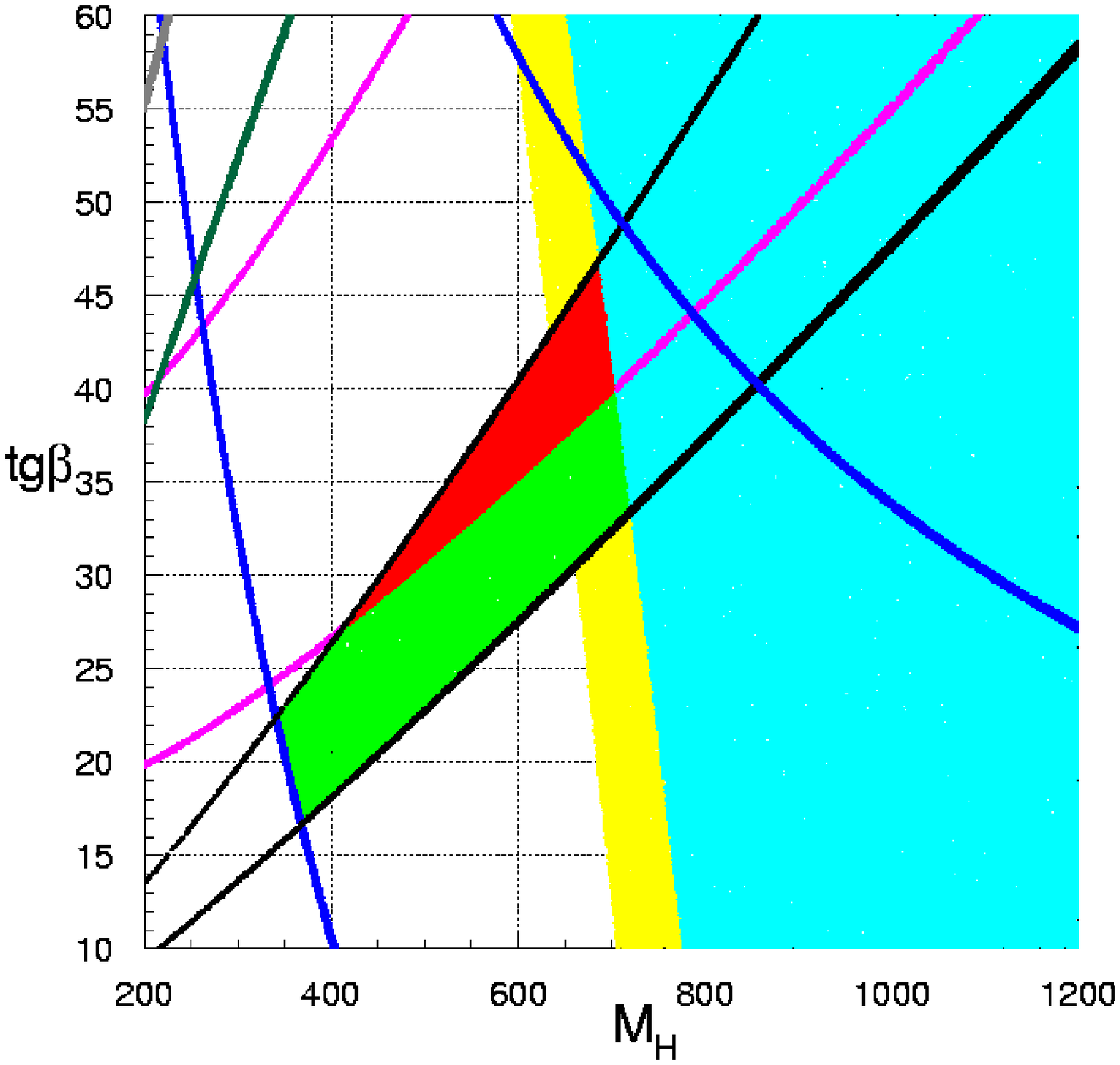}
\includegraphics[scale=0.4]{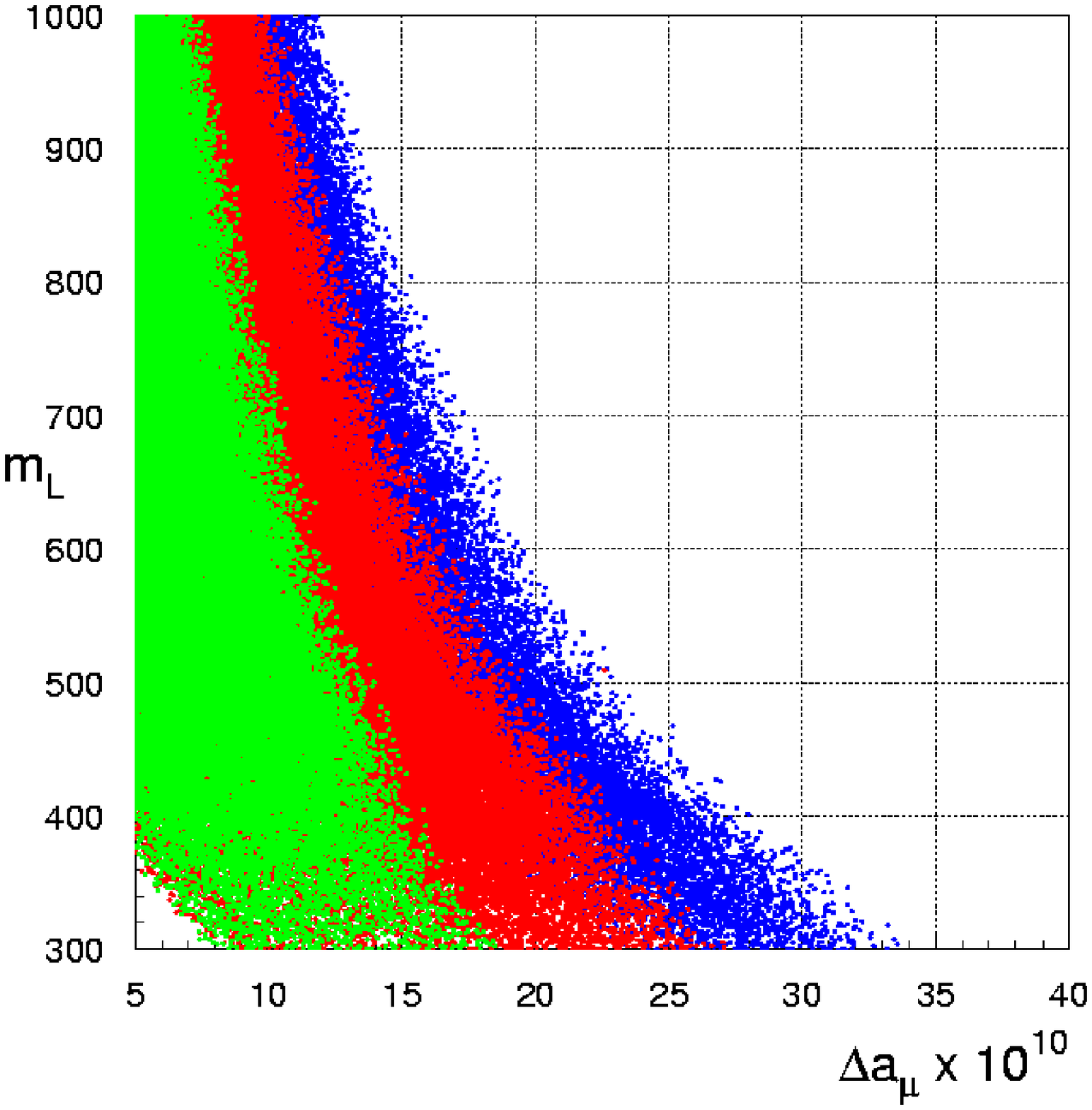}
\caption{\label{fig_MSSM_full}
Left plot: Combined constraints from low-energy observables and dark matter 
in the $\tan\beta$--$M_H$ plane setting $[\mu,M_{\tilde{\ell}}]=[0.5,0.4]$~TeV.
The light-blue area is excluded by the dark-matter conditions \cite{IPwmap}.
Within the red (green) area all the reference values of the low-energy 
observables (but for $a_\mu$) are satisfied.
The yellow band denote the area where the stau coannihilation mechanism 
is active ($1< M_{\tilde{\tau}_R}/M_{\tilde{B}}<1.1$); in this area
the $A$-funnel region (where $M_H\thickapprox 2 M_1$) and the stau 
coannihilation region overlap.
Right plot: $\Delta a_\mu=(g_\mu-g^{\rm SM}_\mu)/2$ vs.~the slepton mass 
within the funnel region taking into account the $B\to X_s \gamma$ 
constraint and setting $R_{B\tau\nu}>0.7$  (blue),
$R_{B\tau\nu}>0.8$  (red), $R_{B\tau\nu}>0.9$ (green) \cite{IPwmap}.
The supersymmetric parameters have been varied in the following ranges: 
200~GeV~$\leq M_{2}\leq$~1000~GeV, 500~GeV~$\leq \mu\leq$~1000~GeV, 
$10\leq \tan\beta \leq 50$.
In both plots, we have set $A_U=-1$~TeV, $M_{\tilde{q}}=1.5~$TeV,
and imposed the GUT relation  $M_1 \approx M_2/2 \approx M_3/6$.
}
\end{figure*}

Within the MSSM, the scenario with large $\tan\beta$ and heavy squarks is
particularly interesting. On the one hand, values of $\tan\beta \sim$ 30--50
can allow the unification of top and bottom Yukawa couplings,
as predicted in well-motivated grand-unified models \cite{GUT}.
On the other hand, a Minimal Flavor Violating (MFV) structure \cite{MFV}
with heavy ($\sim TeV$) soft-breaking terms in the quark sector
and large $\tan\beta \sim 30-50$ values leads to interesting phenomenological
virtues \cite{tgbhints,vives}: the present $(g-2)_\mu$ anomaly
and the upper bound on the Higgs boson mass can be easily accommodated,
while satisfying all the present tight constraints in the electroweak and
flavor sectors. Additional low-energy signatures of this scenario
could possibly show up in the near future in $\cB(\Btaun)$,
$\cB(B_{s,d}\to \ell^+\ell^-)$ and $\cB(B\to X_s \gamma)$.
In the following, as discussed in \cite{IPwmap}, we analyze the above scenario 
under the additional assumption that the relic density of a Bino-like
lightest SUSY particle (LSP) accommodates the observed dark matter distribution
\begin{equation}
0.094 \leq \Omega_{\rm CDM} h^2 \leq 0.129 \quad \rm{at}\: 2\sigma\:
\rm{C.L.}\,.
\label{eq:omg}
\end{equation}
In the regime with large $\tan\beta$ and heavy squarks, the relic-density
constraints can be easily satisfied mainly in the so called $A$-funnel region
\cite{funnel} where $M_{\tilde{B}}\thickapprox M_A/2$.
The combined constraints from low-energy observables and dark matter in the
$\tan\beta$--$M_H$ plane are illustrated in Figure~\ref{fig_MSSM_full} (left).
The light-blue areas are excluded since the stau turns out
to be the LSP, while the yellow band denotes the allowed region where the stau
coannihilation mechanism is also active.
The remaining bands correspond to the following constraints/reference-ranges
from low-energy observables:
\begin{itemize}
\item $B\to X_s \gamma$ [$1.01 < R_{ Bs\gamma} <1.24$]:  
allowed region between the two blue lines.
\item $a_\mu$ [$2 < 10^{-9} (a_{\mu}^{\rm exp} - a_{\mu}^{\rm SM}) < 4$ \cite{gm2}]:
allowed region between the two purple lines.
\item{$ B\to \mu^+ \mu^-$}   [${\cal B}^{\rm exp}< 8.0 \times 10^{-8}$ \cite{Bmm}]: 
allowed region below the dark-green line.
\item{$\Delta M_{B_{s}}$}  [$\Delta M_{B_{s}} = 17.35 \pm 0.25~{\rm ps}^{-1}$ \cite{Dms}]:
allowed region below the gray line.
\item $B\to \tau \nu$ [$0.8 < R_{B\tau\nu} < 0.9$]: 
allowed region between the two black lines [ red (green) area if all the other 
conditions (but for $a_\mu$) are satisfied].
\end{itemize}
From Figure~\ref{fig_MSSM_full} (right), we deduce that there is
a quite strong correlation between $\Delta a_{\mu}$ and $\cB(\Btaun)$
thanks to the $A$-funnel region condition $M_H\thickapprox 2 M_1$.
A SUSY contribution to $a_\mu$ of $\cO(10^{-9})$ generally implies
a sizable effect in $0.7<\cB(\Btaun)<0.9$.
A more precise determination of $\cB(\Btaun)$ is therefore a
key element to test this scenario.
\begin{figure*}[h]
\includegraphics[scale=0.4]{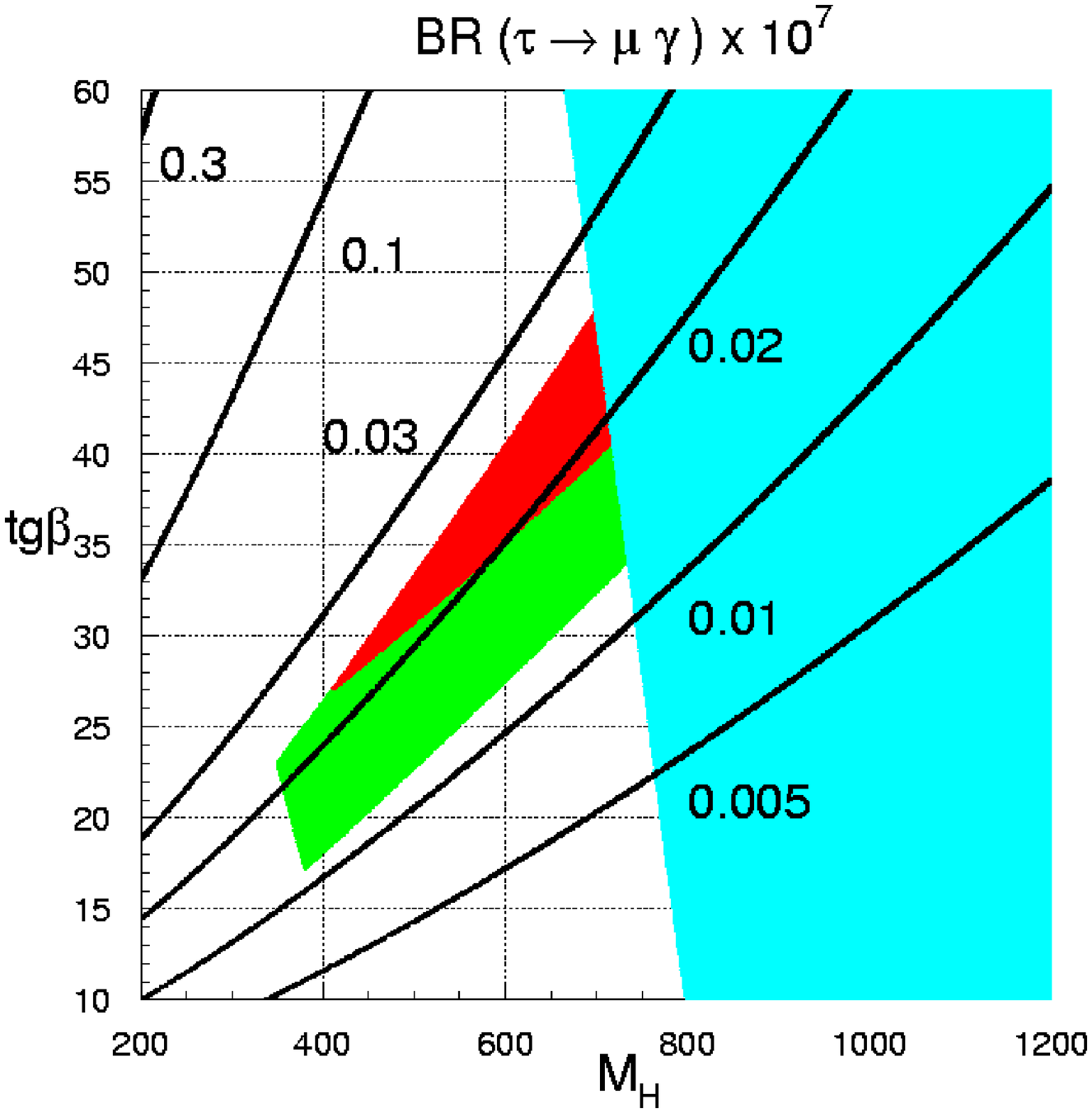}
\includegraphics[scale=0.4]{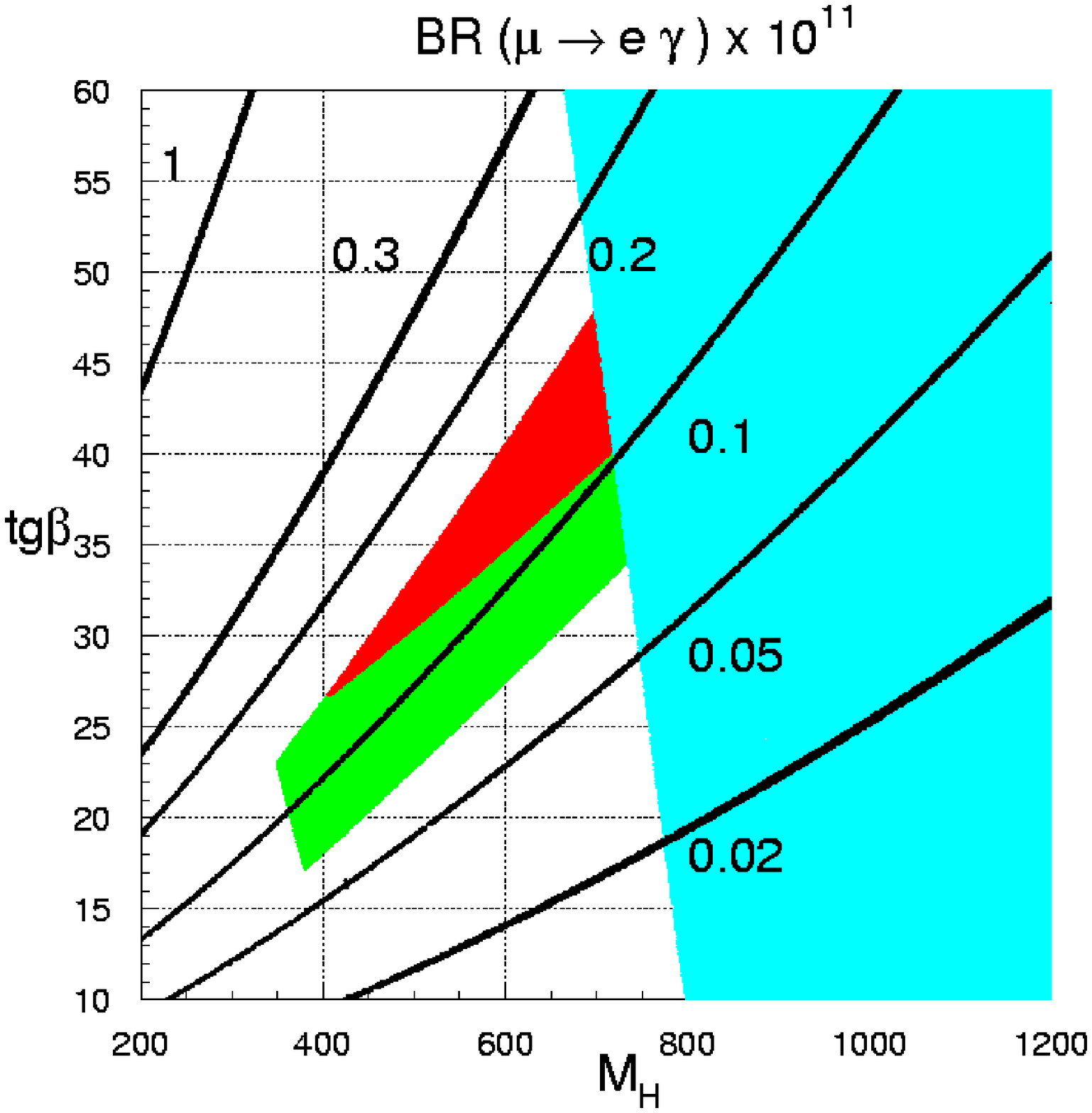}
\caption{\label{fig_meg_full} Isolevel curves for
$\cB(\mu\rightarrow e\gamma)$ and $\cB(\tau\rightarrow \mu\gamma)$
assuming $|\delta_{LL}^{12}|=10^{-4}$ and  $|\delta_{LL}^{23}|=10^{-2}$
in the  $\tan\beta$--$M_H$ plane \cite{IPwmap}. The green/red areas correspond
to the allowed regions for the low-energy observables
illustrated in Figure~\ref{fig_MSSM_full} for
$[\mu,M_{\tilde{\ell}}]=[0.5,0.4]$~TeV.}
\end{figure*}

The interplay of $B$ physics observables, dark-matter constraints, 
$\Delta a_\mu$ of $O(10^{-9})$, and LFV rates is shown in Figure~\ref{fig_meg_full}.
For a natural choice of $|\delta^{12}_{LL}|=10^{-4}$
${\cal B}(\mu\rightarrow e\gamma)$ is in the $10^{-12}$ 
range, i.e.~well within the reach of MEG~\cite{MEG} experiment.
On the other hand, ${\cal B}(\tau\rightarrow \mu\gamma)$ lies within 
the $10^{-9}$ range for a $|\delta^{23}_{LL}|=10^{-2}$, that is
a natural size expected in many models.\\

\begin{acknowledgments}
I wish to thank the conveners of WG3 for the kind invitation and
G.~Isidori, F.~Mescia and D.~Temes for collaborations
on which this talk is partly based.
I also acknowledge support from the EU contract No. MRTN-CT-2006-035482, "FLAVIANET"
and from the Spanish MEC and FEDER FPA2005-01678.

\end{acknowledgments}

\end{document}